\documentclass[twocolumn]{aastex701}

\begin{document}

\title{Flash Ionization of the Early Universe by Pop III.1 Supermassive Stars}

\author[orcid=0000-0002-3389-9142,sname='Tan']{Jonathan C. Tan}
\affiliation{Dept. of Space, Earth \& Environment, Chalmers University of Technology, Gothenburg, Sweden}
\affiliation{Dept. of Astronomy \& Virginia Institute for Theoretical Astronomy, University of Virginia, Charlottesville, VA, USA}
\email[show]{jctan.astro@gmail.com}

\begin{abstract}
The Pop III.1 theory for supermassive black hole (SMBH) formation predicts that a substantial fraction of the early universe was ionized by supermassive stars at redshifts $z\sim20-30$, an era we refer to as ``The Flash''. This is followed by recombination to a mainly neutral state within a few tens of Myr. Here we discuss the implication of this ionization for the scattering optical depth of the cosmic microwave background (CMB), $\tau$. We find a fiducial contribution of $\tau_{\rm PopIII.1}\sim0.04$. Combining this with the contribution to reionization by standard galaxy populations at $z\lesssim 10$ with $\tau_{\rm gal}\simeq0.06$, yields a total of $\tau\simeq0.10$. As noted recently by several authors, such a value may help resolve apparent ``problems'' faced by $\Lambda$CDM of discrepant CMB-based measures of the Hubble constant (``Hubble tension''), as well as negative neutrino masses and dynamical dark energy that have been implied by recent Baryonic Acoustic Oscillation (BAO) results from the Dark Energy Spectroscopic Instrument (DESI). In addition, free-free emission from The Flash boosts the cosmic radio background, which could help explain the large 21-cm absorption depth reported by the Experiment to Detect the Global EoR Signature (EDGES).
\end{abstract}



\section{Introduction}\label{sec:intro}

The Pop III.1 theory for supermassive black hole (SMBH) formation \citep[][]{2019MNRAS.483.3592B,2023MNRAS.525..969S,2025MNRAS.536..851C} \citep[see][for a review]{2024arXiv241201828T} predicts that a substantial fraction of the early universe is flash ionized by rapidly expanding ({\it R-type}) HII regions powered by SMBH-progenitor supermassive stars at redshifts $z\sim20-30$. These stars are born in ``Pop III.1'' dark matter minihalos, i.e., with $\sim10^6\:M_\odot$, which are defined to be the metal-free first-collapsed structures to form in their local region of the universe such that they are not impacted by external feedback, especially ionizing feedback, from astrophysical sources \citep{2008ApJ...681..771M}. Metal-free minihalos that have been ionized or partially ionized, i.e., ``Pop III.2'' sources, are expected to have elevated abundances of $\rm H_2$ and HD, catalyzed by the presence of free electrons, which enhance cooling and thus fragmentation in the minihalo \citep[e.g.,][]{2006MNRAS.373..128G}. The process of weakly interacting massive particle (WIMP) dark matter annihilation in the Pop III.1 protostar, which requires significant adiabatic contraction of the dark matter density in a slowly contracting undisturbed minihalo, can affect the stellar structure \citep[][]{2008PhRvL.100e1101S,2009ApJ...692..574N}, in particular keeping the protostar in a large, relatively cool state \citep[][]{2015ApJ...799..210R,2025arXiv250700870N}, which then may allow it to avoid photoevaporation feedback that typically truncates accretion if there is contraction to the zero age main sequence \citep{2008ApJ...681..771M,2010AIPC.1294...34T,2011Sci...334.1250H,2014ApJ...792...32S,2014ApJ...781...60H}. Thus the protostar may be able to grow to $\sim 10^5\:M_\odot$, with this mass scale set by the baryonic content of the dark matter minihalo, followed by a phase of stellar evolution involving a high production rate of H-ionizing photons. 

Key motivating features of the Pop III.1 model include the prediction that all SMBHs form early in the universe, i.e., by $z\sim20$, as ``heavy'' seeds with $\sim10^5\:M_\odot$ and with the ionizing feedback from pre-SMBH supermassive Pop III.1 stars 
setting the cosmic abundance of SMBHs, with fiducial values of $n_{\rm SMBH}\sim 10^{-1}\:{\rm cMpc^{-3}}$ \citep[][]{2024ApJ...971L..16H,2025arXiv250117675C}. We note that alternative models of heavy seed formation via ``Direct Collapse'' in metal-free irradiated or turbulent atomically-cooled ($\sim 10^8\:M_\odot$) halos struggle to reach this level of abundance by several orders of magnitude \citep[e.g.,][]{2016ApJ...832..134C,2019Natur.566...85W,2022Natur.607...48L,2025arXiv250200574O}. 

Another key feature of the Pop III.1 seeding of SMBHs is that it can naturally explain why there appears to be a characteristic minimum mass scale of the SMBH population, i.e., a dearth of intermediate mass black holes (IMBHs) in the mass range $\sim 10^2-10^4\:M_\odot$ \citep[e.g.,][]{2020ARA&A..58..257G}, or a break in the power law distribution of SMBH masses below $\sim10^6\:M_\odot$ \citep{2024arXiv241017087M}, with this being related to the baryonic mass content of Pop III.1 minihalos. On the other hand, models of SMBH formation via ``light'' seeds, especially via collisional growth in dense star clusters \citep[e.g.,][]{2004ApJ...604..632G,2022MNRAS.512.6192S}, tend to produce many more IMBHs than SMBHs.

As discussed by \citet{2024arXiv241201828T}, a key prediction of the Pop III.1 seeding theory is the presence of ionized bubbles around Pop III.1 supermassive stars at redshifts $z\sim20-30$. Fiducial values of H-ionizing photon luminosities of $S\sim 10^{53}\:{\rm s}^{-1}$ from supermassive stars with WIMP-enhanced lifetimes of $t_*\sim10\:$Myr lead to {\it R-type} expanding HII regions that propagate into the intergalactic medium (IGM) for distances of $R_R\simeq 1.10 t_{*,10}^{1/3}S_{53}^{1/3}\:$cMpc, independent of redshift. These regions that have been flash ionized are then no longer able to form Pop III.1 SMBH progenitors, but rather much lower mass Pop III.2 stars. However, while the particular sizes of the {\it R-type} HII regions impact the overall number density of SMBHs, i.e., given by $n_{\rm SMBH}=3/(4\pi R_R^3)\rightarrow 0.18/(t_{*,10}S_{53})\;{\rm cMpc}^{-3}$, a generic feature of the model is that a large fraction of the universe is ionized, regardless of the detailed properties of the Pop III.1 stars, i.e., ionizing luminosities, lifetimes and thus sizes of their HII regions. Thus our primary goal in this {\it Letter} is to carry out a simple estimate, independent of detailed individual source properties, for the contribution of this predicted early phase of reionization to the scattering optical depth of CMB photons. 

The later phase of reionization powered by ``standard'' galaxy populations has been modeled in many studies \citep[e.g.,][]{2015ApJ...802L..19R,2017MNRAS.465.4838G}, with the general finding that the universe transitions to a mostly ionized state around $z\sim 8$, with the process largely complete by $z\sim 5$. While these models contain a number of uncertainties, especially the assumed values of the initial mass function (IMF) of stars or properties of active galactic nuclei (AGN) that produce the ionizing photons and their escape fractions from their host galaxies to the IGM, they are guided and constrained by observations of galaxy populations. The particular model presented by \citet{2015ApJ...802L..19R} was mainly constrained by data probing the end of the reionization era, but it remains largely consistent with latest {\it James Webb Space Telescope (JWST)} observations out to higher redshifts, $z\sim 10$, with the IGM neutral fraction approaching values close to unity by these redshifts \citep[][R. Ellis, priv. comm.]{2024ApJ...975..208T,2025ApJS..278...33K,2025ApJ...981..134P,2025arXiv250402683M}. 

The ionized IGM produced from these standard galaxy populations is found to contribute a scattering optical depth to CMB photons of $\tau_{\rm gal}\simeq 0.06$, which is consistent with latest published results from the Planck Collaboration \citep{2020A&A...641A...6P}, who found $\tau=0.054\pm0.007$. A more recent analysis by \citet{2021MNRAS.507.1072D} found $\tau=0.063\pm0.005$. Parametric reionization histories that are fit to these CMB data indicate that reionization began at $z\sim10-12$ and ended at $z\sim 5$.

However, a number of recent papers have argued for a larger value of $\tau\simeq 0.09$, i.e., significantly larger than the above CMB-inferred values \citep[e.g.,][]{2025arXiv250305691A,2025arXiv250416932S,2025arXiv250421813J}. Such higher values would help alleviate tensions arising from CMB-based estimates of the Hubble constant (i.e., ``Hubble tension'') and from recent Baryonic Acoustic Oscillation (BAO) measurements from the Dark Energy Spectroscopic Instrument (DESI) 
\citep{2025arXiv250314738D}, 
which, combined with {\it Planck} CMB results, manifest as a preference for negative neutrino masses and evolving, i.e., dynamical, dark energy. However, as has been pointed out by the above authors, the measurement of $\tau$ from the CMB faces a number of challenging systematic uncertainties, i.e., instrumental systematic effects and astrophysical foregrounds, which might yet allow compatibility with a larger value.

If $\tau$ is in fact closer to 0.09, then it has major implications for the reionization history of the universe. In particular, it would require an extra phase of ionization that is not specifically included in most current astrophysical models \citep[e.g.,][]{2015ApJ...802L..19R} \citep[see also][]{2006ApJ...650....7H,2015MNRAS.453.4456V}. While models of so-called ``double reionization'' have been explored previously \citep[e.g.,][]{2003ApJ...586..693W,2003ApJ...591L...5C,2003ApJ...595....1H,2005ApJ...634....1F}, they generally treated an early Pop III phase via simple extensions of normal galactic models, e.g., by varying the IMF (and thus the ionizing efficiency per baryon) and required simple parameterizations for the transition from Pop III to Pop II sources. Furthermore, various studies \citep[e.g.,][]{2003ApJ...595....1H,2005ApJ...634....1F} concluded that a distinct early phase of reionization was unlikely because the transition from Pop III to Pop II would occur in a spatially inhomogeneous and temporally extended manner. 

Given that the observed IGM ionization fraction is near unity at $z\sim 10$ \citep[e.g.,][]{2024ApJ...975..208T}, the results of \citet{2005ApJ...634....1F} indicate that an early phase of increased IGM ionization fraction would require a very distinct high redshift population of ionizing sources. Pop III.1 supermassive stars could be such sources and since an early phase of flash ionization is a key feature and prediction of the Pop III.1 cosmological model of SMBH seeding, we are thus motivated to present a first, simple calculation of its basic properties, including duration and expected contribution to $\tau$.

\section{Contribution to $\tau$ of Pop III.1 Ionization in ``The Flash''}\label{sec:flash}

\begin{figure*}[ht!]
\plotone{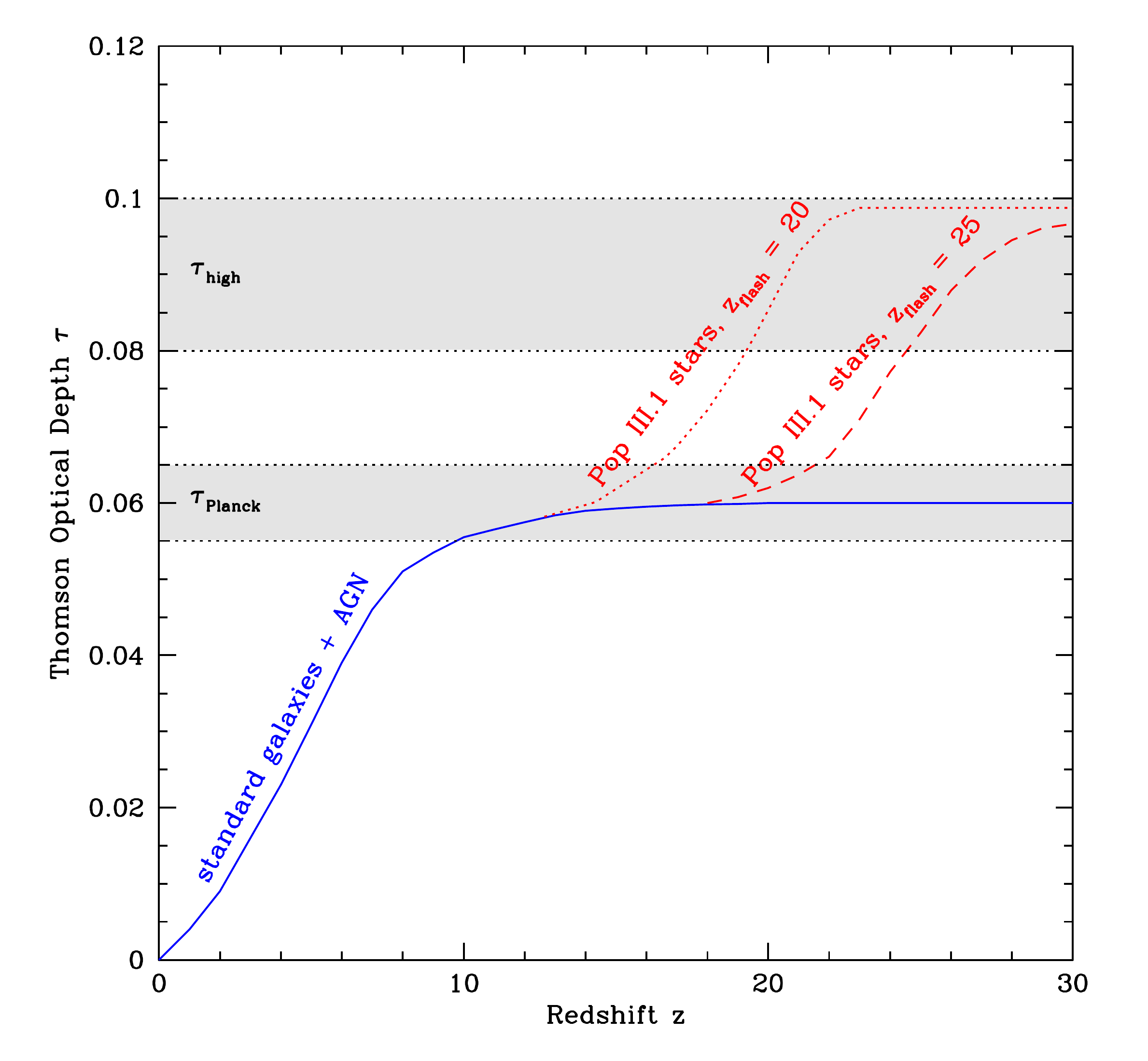}
\caption{Thomson optical depth to electron scattering, $\tau$, integrated out to redshift, $z$. The blue solid line is the estimated contribution from ``standard'' galaxies \citep{2015ApJ...802L..19R} and AGN (although note AGN are expected to make only a minor contribution). The red dotted line shows the contribution from Pop III.1 supermassive stars with epoch of peak flash ionization at $z_{\rm flash}=20$. The red dashed line shows the corresponding model for $z_{\rm flash}=25$ (see text). The lower gray shaded area shows $\tau=0.06\pm0.005$, which is representative of values derived from most recent analyses of {\it Planck} CMB data \citep[e.g.,][]{2020A&A...641A...6P,2021MNRAS.507.1072D}. The upper gray shaded area shows $\tau=0.09\pm0.01$ that is needed to alleviate negative neutrino masses and dynamical dark energy \citep{2025arXiv250416932S,2025arXiv250421813J}.
\label{fig:tau}
}
\end{figure*}

We construct a simple model for reionization by supermassive Pop III.1 stars. We assume these stars all form together at a redshift $z_{\rm form}$, which leads to a subsequent peak level of ionization in their HII regions, $f_{i,{\rm peak}}$, occurring at redshift $z_{\rm flash}$. These HII regions are assumed to fill a significant fraction of the volume of the universe at this epoch, $f_{i,{\rm vol}}$. Based on the cosmological volume SMBH seeding models of \citet{2019MNRAS.483.3592B} and \citet{2023MNRAS.525..969S} that explored and developed the Pop III.1 scenario, we consider two cases: $z_{\rm flash}=20$ and $z_{\rm flash}=25$. Note that these values are constrained by the need to form enough Pop III.1 sources to seed a cosmic population of SMBHs with $n_{\rm SMBH}\sim 10^{-2}-10^{-1}\:{\rm cMpc}^{-3}$ \citep[][]{2024ApJ...971L..16H,2025arXiv250117675C}, requiring isolation distance parameters in the range $d_{\rm iso}\simeq 50-75\:$pkpc, i.e., $\simeq 1-2\:$cMpc at these redshifts. Similar results apply when the isolation distance is set to be equal to the HII region radius around Pop III.1 supermassive stars \citep[][Petkova et al., in prep.]{2024arXiv241201828T,2025arXiv250723004S}.

For our fiducial case we will consider values of $f_{i,{\rm peak}}=0.5$ and $f_{i,{\rm vol}}=0.5$. We note there is simple linear degeneracy between these parameters for the total contribution to $\tau$. However, in the context of the Pop III.1 model, we require $f_{i,{\rm vol}}$ to be near unity. Similarly, HII regions undergoing {\it R-type} evolution are expected to have ionization fractions of order unity when the star is shining, since the mean free path of ionizing photons is relatively short, i.e., $\lambda_{\rm mfp}=(n_{\rm H}\sigma_{\rm p.i.})^{-1} = 9.0 (n_{\rm H}/n_{\rm H,z=30})^{-1}(h\nu/13.6\:{\rm eV})^3\:$pc, where $\sigma_{\rm p.i.}$ is the photoionization cross section of H and we have normalized to the mean IGM density at $z=30$, i.e., $n_{\rm H,z=30}=5.72\times10^{-3}\:{\rm cm}^{-3}$. However, after the star dies and is no longer a significant source of ionizing photons, the ionization fraction drops, even as the {\it R-type} HII region front is still expanding. So the average ionization fraction in the HII region at its maximum extent is expected to be moderately lower than unity.


The timescale for the ionization fraction to rise up to its peak value, $t_{\rm rise}$, is assumed to be intermediate between the lifetime of the ionizing source, i.e., with fiducial value of 10~Myr for the lifetime of a supermassive star that is somewhat prolonged by WIMP annihilation heating \citep[see discussion by][]{2024arXiv241201828T}, and the time to establish a Strömgren sphere, $t_{\rm ion}$, in gas that has a density similar to that of the mean IGM density. The value of $t_{\rm ion}$ assuming mean cosmic density is
\begin{equation}
    t_{\rm ion} =\frac43 \pi R_S^3 \frac{n_{\rm H}}{S}=\frac{1}{\alpha^{(2)}n_{\rm H}}=51.3\left(\frac{1+z_{\rm form}}{31}\right)^{-3}\:{\rm Myr},\label{eq:tion}
\end{equation}
where $R_S$ is the radius of the Strömgren sphere, $n_{\rm H}$ is number density of H nuclei, $\alpha^{(2)}=1.08\times 10^{-13}T_{3e4}^{-0.8}\:{\rm cm^3\:s}^{-1}$ is the recombination rate to excited states of ionized H at a fiducial temperature of $30,000\:$K that is expected in metal-free gas, $S_{53}\equiv S/10^{53}\:{\rm s}^{-1}$, and the final calculation adopts the mean number density of H at $z=30$. Note, also that for simplicity we have ignored the reionization of He. Pop III.1 sources are expected to form in moderately overdense regions. For example, in the simulations of \citet{2025arXiv250723004S} the average density in the HII region around a Pop III.1 supermassive star is a factor of about three times greater than the cosmic average. From such simulations of ionizing feedback it is also seen that when $t_{\rm ion}>t_*$, the {\it R-type} ionization front continues to propagate after the star has ended its life. Given the above considerations we adopt a fiducial value for $t_{\rm rise}=30\:$Myr. We note that this implies Pop III.1 stars would actually have started shining at $z_{\rm form}\simeq 23$ and 30 for our cases of $z_{\rm flash}=20$ and 25, respectively. We also note that $t_{\rm rise}$ can thus additionally be interpreted as encompassing a modest spread in formation redshifts of Pop III.1 stars that is similar in magnitude to these differences.

After reaching a peak ionization fraction in The Flash, we assume this level then decreases exponentially on a timescale equal to the recombination timescale in the HII regions, i.e., given by equation~\ref{eq:tion}, but adopting an overdensity compared to the mean IGM of a factor of three. Note, given the {\it R-type} nature of the HII regions, there is limited impact on the density structure of the gas due to the ionizing feedback. For our cases of $z_{\rm flash}=20$ and 25, these recombination times are 55~Myr and 29~Myr, respectively.

For the above reionization histories we then integrate along a line of sight to evaluate the contribution of each interval to the Thomson optical depth, $\tau$. For this calculation, a helpful reference is that, given the Thomson cross section $\sigma=6.6525\times10^{-25}\:{\rm cm}^2$, the contribution from each co-moving Mpc of mean density IGM that is fully ionized is $d\tau=3.7875\times 10^{-4}\:{\rm cMpc}^{-1}$. Furthermore the co-moving radial distance from $z=20$ to 30 is about 620~cMpc.

Figure~\ref{fig:tau} shows the contributions to $\tau$ from our two example Pop III.1 models with $z_{\rm flash}=20$ and 25. They have been added to the contribution from ``standard'' galaxies and AGN, as calculated by \citet{2015ApJ...802L..19R}. A representative estimate for $\tau$ based on the analysis of {\it Planck} CMB data, i.e., $\tau_{\rm Planck}=0.06\pm0.005$ \citep[][]{2020A&A...641A...6P,2021MNRAS.507.1072D}, is shown by the lower shaded bar. A representative estimate for $\tau$ that is needed to alleviate negative neutrino masses and dynamical dark energy, i.e., $\tau_{\rm high}=0.09\pm0.01$ \citep{2025arXiv250416932S,2025arXiv250421813J}, is shown by the upper shaded bar. We see that the example Pop III.1 models give similar contributions of $\tau_{\rm PopIII.1}\simeq 0.04$, which then yield total values of $\tau\simeq 0.1$. Such optical depths are consistent with the estimates of $\tau_{\rm high}$ that are needed to resolve negative neutrino masses and dynamical dark energy. 

We emphasize that the Pop III.1 theory, which is astrophysically motivated to explain the origin of the entire cosmic population of SMBHs, {\it predicts} that there is an epoch of early flash ionization of the universe \citep{2024arXiv241201828T}. The models presented here have some uncertainties in their choice of $z_{\rm flash}$ and the combination of $f_{i,{\rm peak}}$ and $f_{i,{\rm vol}}$. However, we find that the variation from $z_{\rm flash}=20$ to 30 leads to very minor differences in $\tau_{\rm PopIII.1}$. Furthermore, for $f_{i,{\rm peak}}$ and $f_{i,{\rm vol}}$ we expect values near and bounded by unity, thus motivating our choices of 0.5 for both. For completeness, we note that maximal values of $\tau_{\rm PopIII.1}=0.16$ ($z_{\rm flash}=20$) and 0.15 ($z_{\rm flash}=25$) arise if both $f_{i,{\rm peak}}=1$ and $f_{i,{\rm vol}}=1$.

\section{Conclusions and Discussion}\label{sec:conclusions}

We have presented an estimate for the contribution of Pop III.1 supermassive stars to cosmic reionization as measured by the Thomson optical depth experienced by CMB photons. These Pop III.1 sources, which are invoked as the progenitors of the entire cosmic SMBH population, are predicted to flash ionize a large volume fraction of the universe to high ionization fractions at $z\sim 20-30$, in an era we term ``The Flash''. The duration of The Flash is set, approximately, by recombination of the IGM at these redshifts. The example models that we have presented, including variation of $z_{\rm flash}$ from 20 to 25, yield values of $\tau_{\rm PopIII.1}\simeq 0.04$. When added to the contribution from standard galaxies, $\tau_{\rm gal}\simeq 0.06$, the total optical depth is $\tau\simeq 0.1$. Such a value, while currently larger than allowed by the most recent analyses of {\it Planck} CMB data, has found some favor in its potential ability to resolve problems facing $\Lambda$CDM of the ``Hubble tension'', negative neutrino masses and dynamical dark energy \citep{2025arXiv250305691A,2025arXiv250416932S,2025arXiv250421813J}.

The model we have presented is very simple and highly idealized. It can be improved by detailed studies of individual HII regions around Pop III.1 supermassive stars and semi-analytic cosmic volume simulations that treat individual sources \citep[][Petkova et al., in prep., la Torre et al., in prep.]{2025arXiv250723004S}. In addition, the model should also be developed to include the contributions to reionization from Pop III.2 sources and early phase AGN, which here have been assumed to be negligible. If these sources make a significant contribution to reionization, then this would lead to a more complex, spatiotemporally extended reionization history, which would blur the two distinct phases shown in Figure~\ref{fig:tau}, as well as leading to an overall increase in $\tau$.

The $E-$mode CMB polarization power spectrum has some sensitivity to the redshift dependence of reionization history \citep[e.g.,][]{2017PhRvD..95b3513H,2018A&A...617A..96M}. Future CMB polarization observations, e.g., with LiteBIRD \citep{2023PTEP.2023d2F01L}, are expected to be able to improve on such constraints and thus test the proposed Pop III.1 reionization history. Furthermore, future studies of the CMB with the Simons Observatory \citep{2019JCAP...02..056A} will be able to further test the Pop III.1 prediction of an early phase of flash ionization, especially via observation of the patchy kinematic Sunyeav-Zel'dovich (pkSZ) effect from the peculiar motion of the HII regions around these sources. We note that pkSZ constraints on reionization history, which have been found to be in $2\sigma$ tension with values of $\tau\simeq0.09$ \citep[][]{2025arXiv250515899C}, assume a monotonically decreasing IGM ionization fraction with redshift and so these need to be re-evaluated for the case of the Pop III.1 scenario with a distinct phase of very early flash ionization. We note that the peculiar motions driving the pkSZ signal are reduced at higher redshifts, so the Pop III.1 reionization scenario has the attractive feature of being able to boost $\tau$ in a way that minimizes additional pkSZ contributions.

The signatures of ionized bubbles and/or heating effects on neutral gas around Pop III.1 sources may also be found in surveys of high-$z$ 21-cm emission. Currently, only upper limits on the power spectrum of redshifted 21-cm brightness temperature fluctuations at $z\gtrsim 15$ have been reported. For example, in the redshift range $z=19.8$ to 25.2 limits have been reported from the LOw Frequency ARray (LOFAR) \citep{2013A&A...556A...2V} Low Band Antenna (LBA), with these constraints derived on spatial scales corresponding to wavenumbers of $k\sim 0.038\: h \:{\rm cMpc}^{-1}$ \citep{2019MNRAS.488.4271G}. At slightly lower redshifts, $z=14.2$ to 16.5, upper limits from observations with the Murchison Widefield Array (MWA) \citep{2013PASA...30....7T} have been reported on scales $0.1\: h\:{\rm cMpc}^{-1}\lesssim k \lesssim 1\: h\:{\rm cMpc}^{-1}$ \citep{2021MNRAS.505.4775Y}. Using the New Extension in Nan\c{c}ay Upgrading LOFAR (NenuFAR) \citep{2012sf2a.conf..687Z}, limits have been found on emission from $z=17.0-20.3$ on scales of $k\simeq0.04\:h\:{\rm cMpc}^{-1}$ \citep{2024A&A...681A..62M,2025arXiv250710533M}. However, many of these current limits are thought to be mostly set by systematics, such as the impact of bright foreground sources \citep{2025A&A...696A..56C}. The forthcoming Hydrogen Epoch of Reionization Array (HERA) \citep{2022ApJ...925..221A} is expected to be able to detect 21-cm emission as far as $z\sim 28$ and may be able to determine if there is a characteristic size of early HII regions, expected to be $\sim 1\:$cMpc in the fiducial Pop~III.1 model. Similarly, the Square Kilometer Array (SKA) LOW \citep{2015aska.confE...1K} is expected to be able to directly image neutral hydrogen from scales of arc-minutes to degrees from $z\sim 6$ to 28, allowing detection of the power spectrum of the cosmological signal and the potential to make tomographic images of HII regions \citep[e.g.,][]{2024MNRAS.528.5212B}.

Finally, the Pop III.1 model may be relevant to the sky-averaged (global) 21-cm signal of neutral hydrogen from the early universe. A tentative detection of this signal from $z\sim 13$ to 27 has claimed based on analysis of data from the low-band antenna of the Experiment to Detect the Global EoR Signature (EDGES) \citep{2018Natur.555...67B} \citep[however, see][]{2022NatAs...6..607S}. The signal is centered at $z=17.2$ and features an absorption depth that is at least twice as strong as predicted by standard astrophysical scenarios in $\Lambda$CDM. One potential resolution of this observation involves an enhanced radio background, equivalent to a brightness temperature of 67.2~K, i.e., significantly ($\simeq18\:$K) greater than that of the CMB at these redshifts with $T_{\rm CMB}=49.5\:$K \citep[e.g.,][]{2018ApJ...858L..17F,2019MNRAS.486.1763F}.

Free-free emission from The Flash could make a contribution to such a background. The brightness temperature arising from path integrated free-free emission is $T_{B,{\rm ff}} = 1124 \nu_{\rm 1.4GHz}^{-0.118} T_{3e4}^{-0.323} [EM/({\rm cm^{-6}pMpc}]\:{\rm K}$, where the emission measure is defined as $EM\equiv\int n_e n_p ds$. Evaluating $T_{B,{\rm ff}}$ with respect to average IGM densities at $z=25$, for emission weighted overdensity clumping factor of $f_{\rm clump}=10$, and integrating over a path length of 10~pMpc, yields: $T_{B,{\rm ff}} = 12.8 \nu_{\rm 1.4GHz}^{-0.118} T_{3e4}^{-0.323} f_{\rm clump,10}^2 (s/{\rm 10\:pMpc}) [(1+z)/26]^6\:{\rm K}$. Integrating this free-free emission through the reionization histories described in \S\ref{sec:flash} yields $T_{B,{\rm ff}} = 0.91$ and $2.9\:$K for $z_{\rm flash}=20$ and 25, respectively. These values would double if we set $f_{i, {\rm peak}}=1.0$ and $f_{i, {\rm vol}}=0.25$, which keeps the same value of Thomson optical depth. Thus, given the sensitivity of the integrated free-free emission to HII region number density, i.e., $\propto f_{\rm clump}^2(1+z_{\rm flash})^6$, this process could lead to a significant radio background that may help in the interpretation of the reported absorption signal from EDGES. Conversely, confirmation of the EDGES 21-cm absorption depth would place more stringent constraints on the Pop III.1 model, in particular favoring relatively early formation redshifts that result in higher density HII regions. This further motivates additional exploration of the Pop III.1 flash reionization scenario, including its predicted free-free background, via more advanced numerical simulations.

\begin{acknowledgments}
I thank, in particular, Richard Ellis and Frederick Davies for stimulating discussions, along with other participants in the SMBH-2025 workshop, where the ideas for this study were germinated. I thank the anonymous referees for helpful comments. I also thank Nilanjan Banik, Christopher Cain, Vieri Cammelli, Bruce Draine, Xiaohui Fan, Katherine Freese, Ivelin Georgiev, Zolt\'an Haiman, Matthew Hayes, Colin Hill, Cosmin Ilie, Ben Keller, Matteo la Torre, Chris McKee, Pierluigi Monaco, Andrew Mummery, Devesh Nandal, Aravind Natarajan, Maya Petkova, Mahsa Sanati, Jasbir Singh, David Spergel, Konstantinos Topalakis, John Wise, Naoki Yoshida \& Alice Young, among others, for fruitful collaborations and discussions related to the Pop III.1 project. I also thank Robert Lupton and his collaborators for developing the SuperMongo graphics software and Ned Wright for developing his Java Script Cosmology Calculator \citep{2006PASP..118.1711W}. I acknowledge funding from ERC Advanced Grant MSTAR (788829), the Chalmers Initiative on Cosmic Origins (CICO), the Virginia Initiative on Cosmic Origins (VICO), and the Virginia Institute for Theoretical Astrophysics (VITA), supported by the College and Graduate School of Arts and Sciences at the University of Virginia.
\end{acknowledgments}

\bibliography{PopIII.1-tau}{}
\bibliographystyle{aasjournalv7}



\end{document}